# NANOINDENTATION-INDUCED DEFORMATION TWINNING IN

# MAX PHASE Ti$_2$AlN

Christophe TROMAS[1], Salomé PARENT[1], Wilgens SYLVAIN[1], Ludovic THILLY[1], Gilles

RENOU[2], Christopher ZEHNDER[3], Sebastian SCHRÖDERS[3], Sandra KORTE-KERZEL[3], Anne

JOULAIN[1]

[1] Intitut Pprime, UPR 3346 CNRS - UNIVERSITE de POITIERS – ENSMA, Département de
Physique et Mécanique des Matériaux, 11 Bd Marie et Pierre Curie, Site du
Futuroscope, TSA 41123, 86073 Poitiers CEDEX 9, France

[2] SIMaP-PM, 1130 Rue de la Piscine, Batiment ECOMARCH, BP75, 38402 St Martin
d'Heres cedex, France

[3] Institute of Physical Metallurgy and Materials Physics, RWTH Aachen University,
52056 Aachen, Germany

Corresponding author: christophe.tromas@univ-poitiers.fr   +33 5 49 49 66 60

## Abstract

Plastic deformation mechanisms have been investigated in the MAX phase Ti$_2$AlN.

Nanoindentation has been used to induce plastic deformation in a single grain, and a

Transmission Electron Microscopy (TEM) lamella has been extracted in cross section

through the indent by using Focused Ion Beam (FIB) technique. By combining TEM

observations and automated crystal orientation mapping (ACOM), highly misoriented

domains (HMD) have been revealed below a nanoindentation imprint. Thanks to a

careful analysis of the relative crystal orientations between these HMD, $\{11\overline{2}2\}$ and

$\{11\overline{2}1\}$ deformation twins have been identified for the first time in a MAX phase.

Complex structures, involving secondary twinning or different $\{11\overline{2}2\}$ twin variants have also been characterized.

## Keywords :

## 1. Introduction

MAX phases are a class of ternary nitrides or carbides [1–4]. Their specific properties result from their nanolaminated structure, alternating metal atomic layers and carbide or nitride layers. Concerning their mechanical properties, they present a rather brittle behavior at room temperature even if at the same time they are damage tolerant [5]. Their mechanical properties evolve toward a more ductile behavior at high temperature [6,7]. Since they possess a hexagonal structure with a high c/a ratio, their plastic deformation at room temperature is often observed to be ruled by basal plane dislocations: most of the transmission electron microscopy (TEM) observations reported in the literature describe perfect dislocations gliding in the basal plane with a Burgers vector $\vec{b} = 1/3 \langle 11\overline{2}0 \rangle$ [8–14], and forming pile ups or walls. However, as observed in many hexagonal metals, basal slip is not sufficient to account for arbitrary deformation, so that when the Schmid factor on basal systems is low, the crystal must find a way to reorient in order to make basal slip more favorable. For MAX phases, one way consists in developing kink bands, which are reoriented regions surrounded by two basal dislocation walls of opposite Burgers vectors acting as low angle tilt boundaries. Due to geometrical incompatibilities, these kink bands are associated with the formation of delamination cracks at their basis when they grow [15,16]. Kink bands have also been reported in many hexagonal metals like cadmium and zinc [17,18]. However, in hexagonal metals, deformation twinning is also a common deformation process allowing for local reorientation [19]. Indeed, a deformation twin is a region of a crystal that has undergone a simple shear in such a way that the resulting structure is identical to that of the parent (matrix), but differently oriented. For a complete description of deformation twinning, the reader is referred to the books of Christian [20] and Kelly and Goves [21] or to the papers of Partridge [22] and Christian and Mahajan [23]. In hexagonal metals, twinning is mostly observed for $\{10\bar{1}2\}$, $\{10\bar{1}1\}$, $\{10\bar{1}3\}$, $\{11\bar{2}1\}$, $\{11\bar{2}2\}$, $\{11\bar{2}3\}$ and $\{11\bar{2}4\}$ planes. Furthermore, if slip is generally easier than twinning in metals, this is the opposite for many ceramic materials[24].

Deformation twinning has been ruled out from the very first founding papers on plastic deformation in MAX phases [7,15,25] mainly because of the Hess and Barrett paper [18] from 1949 where it was said that *"kink bands are expected only in those crystals that are not subjected to twinning"*. However, since 1950, kink bands and deformation twins have been identified as complementary deformation processes in different hexagonal metals like zinc or magnesium [23,26–29], since kink bands allow to accommodate the shear strain due to twinning when the twin is not propagated through the whole crystal. More recently, Wada *et al.* [16] performed TEM analysis around nanoindentations in the MAX phase $Ti_2AlC$. They observed kink bands, but also deformation bands with such a high misorientation angle that they were supposed to be mechanical twins, even if the twinning system could not be identified. However, even in recent reviews [30] describing the many defects reported in MAX phases, twinning is still not considered.

The objective of this paper is thus to investigate the plastic deformation mechanisms involved in the Ti$_2$AlN MAX phase, paying particular attention to deformation twinning. For this purpose, we have used Automated Crystal Orientation Mapping (ACOM) to combine conventional Transmission Electron Microscopy (TEM) observation and crystallographic orientation maps in an indented Ti$_2$AlN MAX phase sample to reveal the existence of $\{11\overline{2}2\}$ and $\{11\overline{2}1\}$ deformation twinning.

## 2. Material and methods

### 2.1. Sample preparation

The Ti$_2$AlN polycrystalline sample was prepared by powder metallurgy using the hot isostatic pressing (HIP) technique with a rise in temperature during 45 min to reach 1450°C. The temperature was maintained over 480 min and finally, the sample was cooled down to room temperature in 480 min. Then the sample was first polished with diamond suspensions and then chemo-mechanically polished with a colloidal alumina suspension.

### 2.2. Nanoindentation

Nanoindentation mechanical testing has been used in order to probe single grains. This configuration is expected to make the analysis of the deformation structure easier and has been used already to investigate plasticity in MAX phases [10,16,31–34].

Nanoindentation tests were performed with a "NanoTest Platform 3" from Micro Materials [35] equipped with sapphire Berkovich tip supplied by Synton. Since the initial purpose was to investigate the brittle to ductile transition (BDT) in $Ti_2AlN$, the nanoindentation tests have been performed at 800°C under an 8.6 $10^{-6}$ mbar vacuum, the BDT temperature for $Ti_2AlN$ being about 900°C [36]. It must be mentioned that for these experiments, the sapphire tip was not heated, so that, due to heat loss in the contact area, the temperature in the indented region was more probably close to 700°C. The indents were characterized by Atomic Force Microscopy (AFM) in tapping mode using a Dimension 3100 microscope from Bruker. Images were processed with the WSxM software [37].

## 2.3. Transmission electron microscopy

A Transmission Electron Microscope (TEM) thin foil was extracted across an indent by focused ion beam (FIB) with a Helios Nanolab 600i, FEI Inc. The lamella was cut perpendicular to the basal plane, and was about 12 µm long, 10 µm wide and 80 to 100 nm thick. The lamella was then analyzed by TEM with a CM20 microscope from Philips. However, due to the high strain induced by a nanoindentation test, it can be difficult to fully characterize the deformation structure by conventional transmission electron microscopy. Automated Crystal Orientation Mapping (ACOM-TEM) allows establishing crystallographic orientation map in the TEM lamella with a spatial resolution better than 10 nm and with an angular resolution close to that provided by Electron Back Scattering Diffraction (EBSD). This technique can thus be used to analyze local reorientation at small scales. Furthermore, there have been many studies during the last 20 years using crystallographic orientation image mapping techniques, such as EBSD, to identify twinning in different materials [38–44], and which provide now a well-established protocol for twinning investigation at small scales. Crystallographic orientation maps were thus established on this lamella thanks to the ACOM-ASTAR technique in a Jeol 2100F TEM. The orientation data were analyzed with different softwares: ATOM [45], the MTEX-Matlab toolbox [46] and Pycotem [47].

## 3. Results

### 3.1.    TEM characterization of dislocations

Fig. 1 presents a 100 mN indent, with a residual depth of 300 nm, obtained at 800°C with a Berkovich indenter in a polycristalline $Ti_2AlN$ sample. Figure 1a combines an AFM surface observation of the residual indent, and TEM image obtained in cross section across the indent, the indent axis being contained in the TEM lamella. It can be observed that the indent is localized in a single grain (grain 1), but close to a grain boundary (grain 2). In the AFM image, few slip lines are observed. Their orientation is consistent with basal slip, and they have been used to choose the orientation of the TEM lamella perpendicular to the basal plane orientation in grain 1. The TEM image in Fig 1a is a bright field image of the whole lamella and has been obtained without any tilt of the lamella in the microscope. Such a condition allows a real correspondence between the observations at the surface by AFM and in the volume by TEM. Fig. 1b and 1c present two TEM images, obtained in different diffraction conditions of the indented area. Fig.

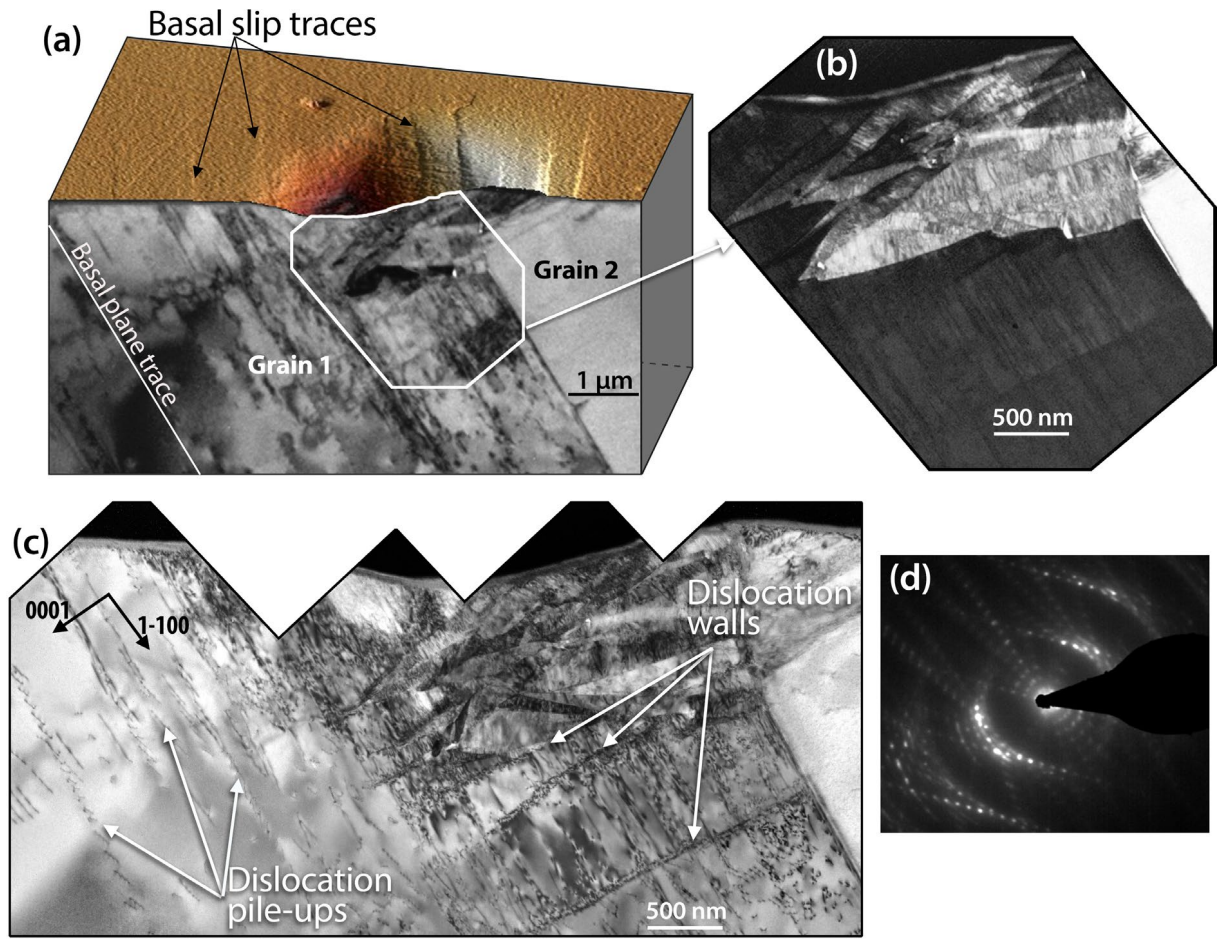

*Figure 1 : (a) Atomic Force Microscopy (AFM) image of a 100 mN indent in Ti₂AlN combined with a Transmission Electron Microscopy (TEM) image of the underlying microstructure observed in cross section. (b) Highly misoriented domains revealed by TEM below the indent and (c) larger view of the microstructure below the indent showing the highly misoriented domains as well as dislocation walls and dislocations pile-ups. (d) Diffraction pattern obtained from different misoriented domains.*

1b is a bright field image where the grain 1 is in zone axis, the electron being parallel to the $[11\bar{2}0]$ direction. The lamella has been tilted by about 20° in the microscope to reach this orientation. In this condition, grain 1 presents a dark contrast while the area below the indent presents a bright contrast indicating disorientation relative to the grain. The basal plane is edge on in grain 1, and the basal plane trace is indicated in figure 1a by a white line. The TEM image in figure 1c is a bright field image obtained with $g = [1\bar{1}00]$. It reveals a complex deformation structure below the indent, composed of lenticular or triangular shape domains, associated to local crystallographic misorientations as confirmed by the diffraction pattern (Fig. 1d). This diffraction pattern shows that these domains are all oriented with a common $[11\bar{2}0]$ direction, parallel to the electron beam, but with a different $[0001]$ orientation. Below and around the highly misoriented regions, basal plane dislocations are observed organized in the classical MAX phase configurations: pile-ups in the basal plane or walls perpendicular to the basal plane.

Fig. 2a presents a misorientation map obtained by the ACOM ASTAR technique in the same TEM lamella

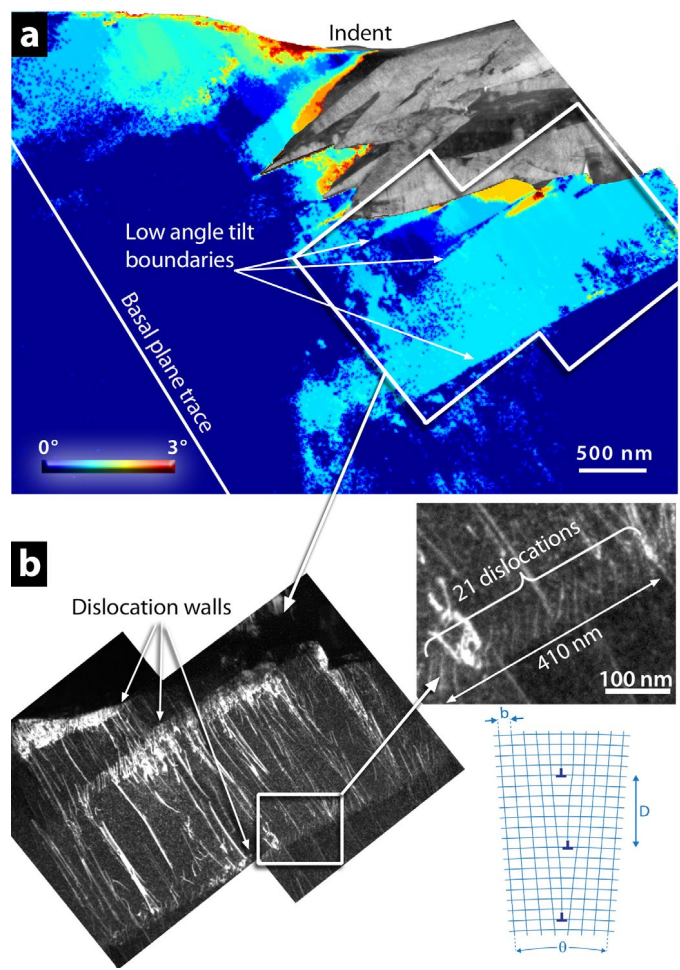

Figure 2 : (a) Misorientation map relative to the non-deformed matrix (angular range 0°-3°) and superposition of the Transmission Electron Microscopy (TEM) image of the highly misoriented domains. (b) Weak beam dark field TEM image of the region highlighted in (a). Dislocation walls are observed and correspond to the low angle tilt boundaries identified in (a)

in grain 1, below the indent. This map represents, in a 0-3° color scale, the local crystallographic misorientation relative to the undeformed region, which is the lower left corner of the image. With this low angular range, the highly deformed zone below the indent is saturated, this is why the TEM image has been superimposed to the misorientation map. The objective is here to focus on the low angle boundaries that clearly appear (in the region surrounded by a white line), at a depth higher than 1.5 μm from the indented surface, which is more than 5 times the depth of the residual indent, and where only little deformation is expected.

In the misorientation map shown in Fig. 2a, low angle tilt boundaries are identified, perpendicularly to the basal traces. They correspond to a 1° misorientation according to the ACOM ASTAR data. The same area has been studied by TEM in weak beam (WB) dark field, as shown in Fig. 2b. In this case, the WB image is obtained with $g = [\bar{2}110]$, close to the $g = [01\bar{1}0]$ zone axis. In this orientation, the basal plane is edge on, and all dislocations are in contrast with the used g vector. Many dislocations, lying in the basal plane, are observed. They form dislocation walls perfectly aligned with the low angle misorientation boundaries visible in the misorientation map in Fig. 2a. This dislocation configuration is expected for MAX phases and corresponds to the classically observed low angle tilt boundaries. The light blue band observed in Fig. 2a is a kink band, composed of two dislocation walls leading to two boundaries of opposite misorientation angle. A closer analysis, as presented in the magnification of Fig. 2b, allows determining the dislocation density in these walls. As an example, 21 dislocations are observed over a 420 nm length of the wall. Considering the lattice parameters of $Ti_2AlN$, this dislocation density (assuming that they all have the same Burgers vector) leads to a theoretical tilt of 0.9°, which is in very good agreement with the 1° ASTAR measurement (see color scale in Fig. 2a).

## 3.2.   Observation of highly misoriented domains

Fig. 3 presents an ACOM ASTAR analysis of the crystallographic misorientations underneath the indent, where the highly deformed zones have been revealed by TEM in Fig. 1. Fig. 3a is an ASTAR map that represents the local misorientation angle relative to the non-deformed region (lower left corner) as in Fig. 2a, but with a wide scale ranging from 0° to 70°. The traces of the basal plane, determined from the ASTAR measurement, are also indicated as thin black lines to make it easier to understand the local configuration. This map shows highly misoriented domains (HMD) that can be considered as subgrains generated during the indentation process. These HMD have been numbered from 1 to 26 for an easier description, number 1 corresponding to the reference matrix in grain 1. According to this representation, the HMD present a large spectrum of misorientations relative to grain 1, this misorientation reaching values as high as 70° for grains 5 and 14. Inside some HMD, local misorientations are observed as in HMD 2, 3, 4 or 24. The highest local misorientations between adjacent HMD have also been reported in large red arrows in Fig. 3a. Local misorientation as high as 72° between HMD 2 and 3 or 80° between HMD 3 and 5 can be observed. It must be also emphasized that there is no misorientation gradient across the boundaries between these HMD. For each subdomain, the basal plane is close to an edge on orientation, as observed in the (0001) pole figure presented in the inset in Fig. 3a. This pole figure is consistent with the TEM diffraction pattern shown in Fig 1b. All the HMD can thus be understood, for symmetry reasons, as local rotations of the crystal around an axis always roughly perpendicular to the extracted TEM lamella.

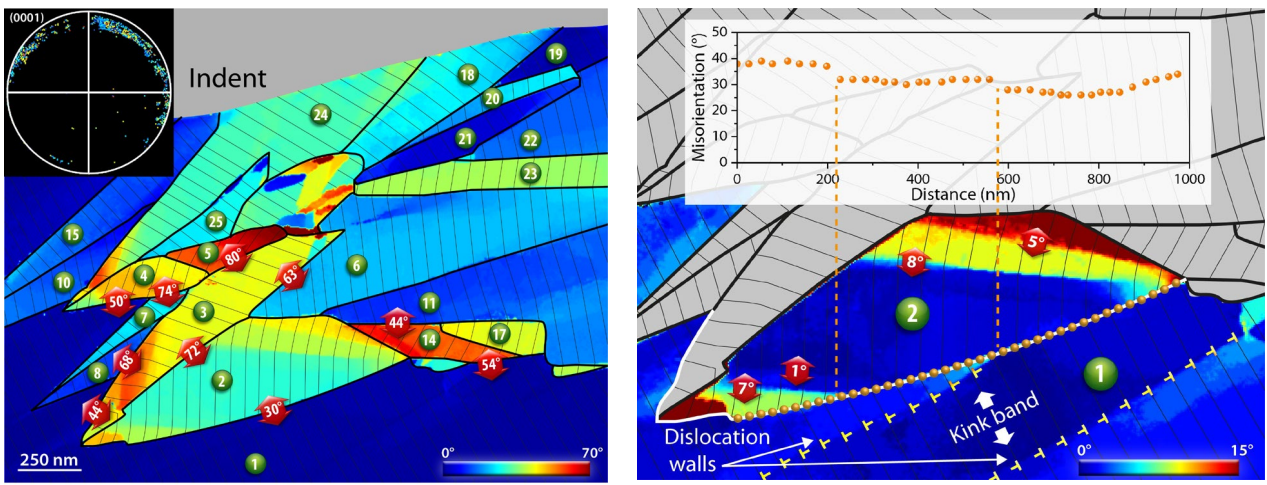

*Figure 3 : (a) Crystallographic misorientation map obtained by the Automated Crystal Orientation Mapping (ACOM) ASTAR technique (color scale range: 0° - 70°). The misorientation angles are relative to the non-deformed region, in the lower left corner. The red arrows indicate the local misorientation between neighbor domains. The basal plane traces have been indicated by dark line. In the inset, (0001) pole figure over the whole region. (b) Local crystallographic misorientation map for domains 1 and 2 (color scale range 0° - 15°). The angles are relative to the mean orientation for each domain. The plot in the inset presents the evolution of the misorientation angle between domains 1 and 2 along their common boundary.*

In some of this HMD, like HMD 2, 3, 4 or 5, some internal low angle misorientations are also observed. In order to emphasize these internal misorientations, figure 3b presents a local misorientation map for the matrix 1 and the HMD 2 where the misorientation is represented for each domain relative to the mean domain orientation, with a narrow scale ranging from 0° to 15°. For the matrix 1, the low angle tilt boundaries already observed in figure 2a are still visible and have been highlighted by a schematic representation of the dislocation walls. In the case of HMD 2, internal low angle tilt boundaries, perpendicular to the basal plane traces, are clearly highlighted. The red arrows indicate the local misorientations across these tilt boundaries, ranging from 1° to 8°. The central dark blue band in HMD 2 is surrounded on one side by a 8° tilt boundary, and on the other side by two close tilt boundaries of 1° and 7°. This band is likely to be a kink band like the one observed in the matrix 1 and the low angle tilt boundaries are likely to be dislocation walls. Furthermore, the local misorientation between the matrix 1 and HMD 2 has been plotted all along the boundary between these two domains. It can be observed that this misorientation is not constant, and ranges from 27° to 38° with two steps of few degrees. These two steps appear at the intersection of the HMD boundary with the previously described low angle tilt boundaries. In fact, the upper part of this HMD boundary has probably absorbed the upper dislocation wall of the kink band identified in the matrix 1, leading thus to a modification of the local misorientation between the matrix 1 and the HMD 2.

These observations evidence a rather complex deformation structure with HMD and dislocation walls forming kink bands inside these HMD. However, if dislocation walls and kink bands are often described as the main plastic deformation mechanism in MAX phases, they cannot explain the very high misorientation here observed between the HMD. If one assumes a tilt boundary with the highest dislocation density, that is one dislocation every c-lattice parameter, the tilt angle would be less than 25°, which is far from the 72° or 80° observed between HMD 2-3 and HMD 3-5 for example. If these HMD were not kink bands, thus the high and abrupt misorientations, along lenticular or needle shape domains, would inevitably suggest deformation twinning. In order to assess this hypothesis, the crystallographic relationships between neighboring HMD will be examined.

## 4. Analysis and discussion

### 4.1. Identification of $\{11\bar{2}2\}$ twin boundaries

To go further with this assumption of deformation twinning in $Ti_2AlN$, the following analysis will be focused on the less deformed region, which corresponds here to the HMD 6 and 18 to 23 shown in figure 3a. It must be reminded that nanoindentation is a dynamic deformation process, where the plastic zone expands all along the test, with regions evolving from a tensile to a compressive stress state. The region along the indent axis has thus accumulated plastic deformation all along the indentation process, while the region close to the edge of the indent has been plastically deformed only in the last stage of the indentation process. This last region is thus expected to present a more simple deformation structure likely to make easier the identification of the elementary deformation mechanisms.

In this part, we will demonstrate the existence in this region of several $\{11\overline{2}2\}$ deformation twins by a careful analysis of the crystallographic orientation maps established by ACOM ASTAR technique. The crystallographic relationship between the parent lattice and the twin lattice is described by four independent twinning elements $K_1$, $K_2$, $\eta_1$ and $\eta_2$ [20,48,49]. $K_1$ is the twinning (or composition) plane, which is an invariant plane of twinning shear, $K_2$ is the undistorted (but rotated) plane called conjugate twinning plane, $\eta_1$ is the direction of shear (or twinning direction) and $\eta_2$ is the conjugate shear direction.

The method to identify twin relationships in crystallographic orientation maps, as proposed in 2002 by Wright and Larsen [38] and Mason *et al*. [39] consists of checking three criteria across a potential twin boundary:

- The twinned region, which can be derived geometrically by a reflection of the parent crystal at the twin plane $K_1$, must satisfy a specific misorientation relationship with the parent crystal. For a $\{11\overline{2}2\}$ twin in Ti$_2$AlN, this misorientation is 24.82° (cf. figure 4).

- The two crystal lattices, on each side of the boundary, must share a common $K_1$ normal.

- The boundary plane separating the two lattices must coincide with a particular twinning plane, so that the trace of the boundary must correspond to the $K_1$ plane trace.

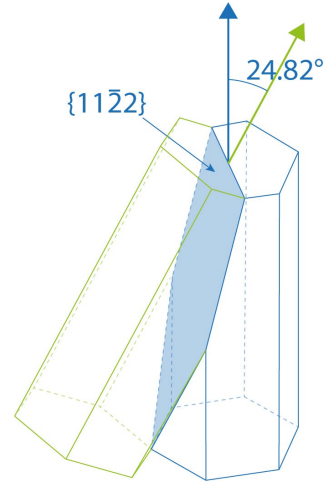

Figure 4 : $\{11\overline{2}2\}$ reflection twin for a hexagonal crystal with c/a=4.55. The misorientation angle between the parent and the twin is 24.82°.

Figure 5 presents such an analysis on five HMD boundaries. The figure is a magnification of the misorientation map presented in Fig. 3a, with the color corresponding to the misorientation angle calculated relative to the non-deformed region (scale ranging from 0° to 70°). The local misorientation across the boundaries is also indicated by orange arrows. For the five boundaries, this local misorientation is very close to the expected 24.82°. Furthermore, for each boundary, the $\{11\overline{2}2\}$ pole figures of the neighboring HMD pairs have been plotted, projected in the X-Y lamella surface. The color code for the poles is relative to the HMD label. For each HMD, the pole figures shows orientation spread of few degrees, probably because of internal dislocation plasticity [41]. This is not surprising considering the highly heterogeneous nature of the stress field generated during the nanoindentation test. For each boundary, it can be observed that there is always a $\{11\overline{2}2\}$ pole from one HMD overlapping another one from the neighbor HMD.

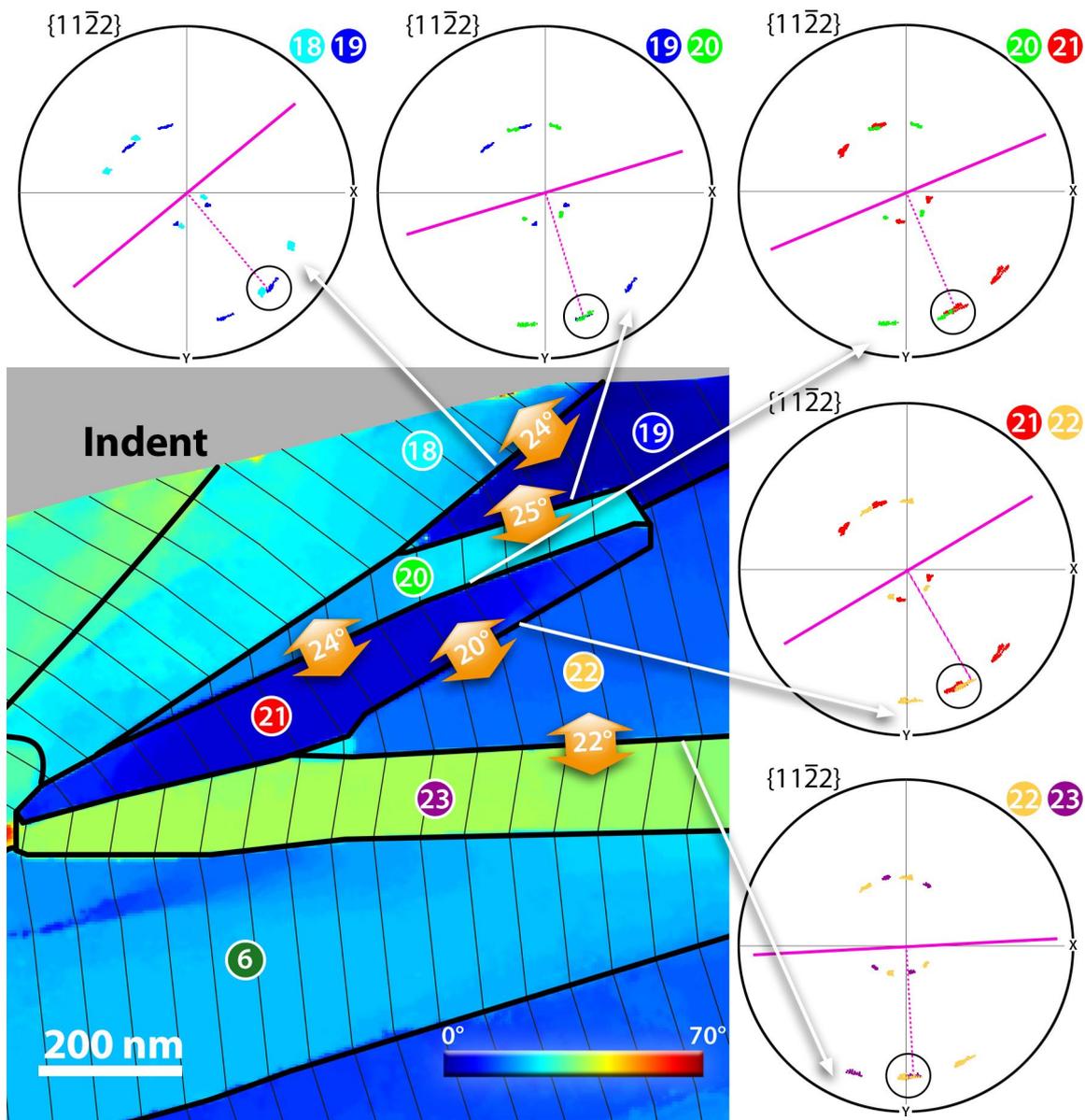

*Figure 5 : Misorientation map relative to the non-deformed matrix (angular scale 0°-70°). The relative misorientation between neighboring HMD is indicated in orange arrows, and the $\{11\overline{2}2\}$ pole figures of the HMD pairs have been plotted. The circles in the pole figures highlight the overlapping of a pole from two neighbor HMD, and the pink solid lines indicate the direction of the HMD boundaries transferred to the pole figures.*

This demonstrates that for the five tested boundaries, the HMD pairs are in such a relative orientation that they share a common $\{11\overline{2}2\}$ plane. For each pole figure, the trace of the boundary on the map has also been transferred in a pink solid line, as well as the normal to this direction in pink dotted line [50]. For each case, this trace is in very good agreement with the $\{11\overline{2}2\}$ plane trace corresponding to the overlapped poles. Since the three criteria are verified, it can be concluded that the five probed HMD boundaries present $\{11\overline{2}2\}$ twin relationship.

To go further, mechanical twins are generally considered as reflection twins [51]. In this case, the two twin-related crystals present an orientation relationship that can be described by a reflection across $K_1$, as well as by a 180° rotation about the direction normal to $K_1$. This property has been used as a definitive check of the existence of $\{11\overline{2}2\}$ deformation twins in Ti$_2$AlN. Since a $(\overline{1}2\overline{1}2)$ twin boundary has been identified between the HMD 22 and 23, a stereographic projection, perpendicular to the surface, has been plotted for these two HMD (cf. Figure 6). A 180° rotation around the normal to $\{\overline{1}2\overline{1}2\}$ has been applied to the stereographic projection of the HMD 23. The result fits perfectly with the stereographic projection of the HMD 22, which lead to the unambiguous conclusion that HMD 22 and 23 have relative crystallographic orientations that satisfy the definition of $\{11\bar{2}2\}$ twinning.

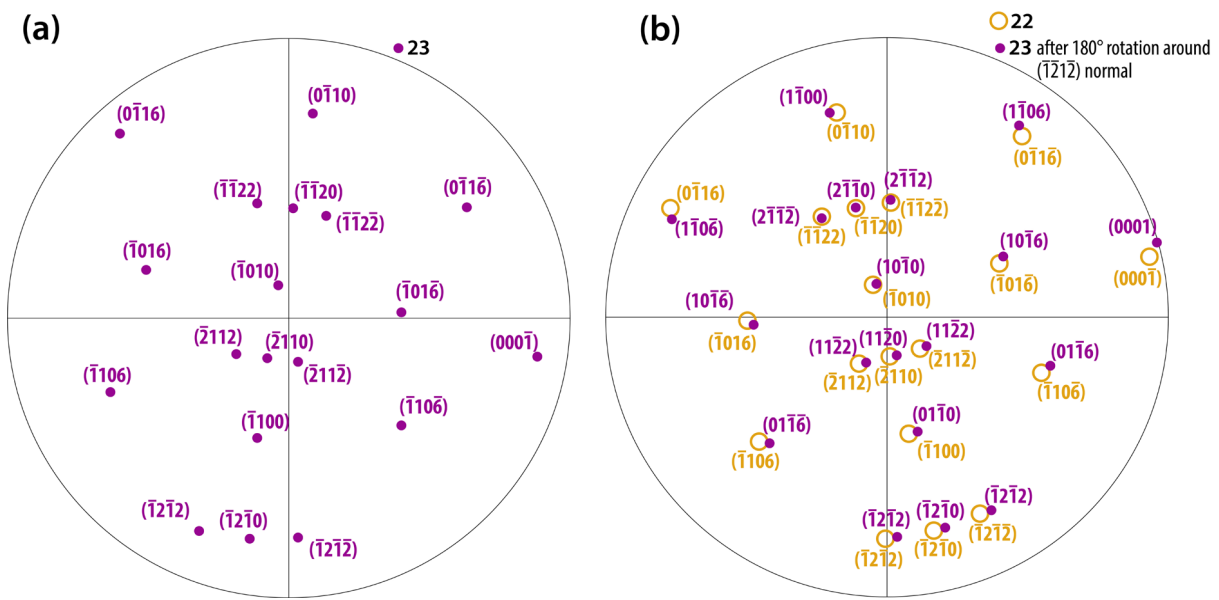

*Figure 6 : (a) stereographic projection, perpendicular to the surface, of HMD 23. (b) Stereographic projection of HMD 23 after a 180° rotation around the normal to $(\overline{1}2\overline{1}2)$ (in purple) and stereographic projection of HMD 22 (in yellow). The superposition of these two projections demonstrates the $(\overline{1}2\overline{1}2)$ twin relation between HMD 22 and 23.*

## 4.2. Scenario for the mechanical twin formation

The deformation structure below the indent is complex with many HMD. We have chosen here to focus the analysis on the simplest region, which is the one that has been deformed later during the indentation test so that it has cumulated less deformation. In the previous paragraph we have analyzed the HMD boundaries to identify $\{11\bar{2}2\}$ twin boundaries. The question that arises from the identification of these twin boundaries is that of the chronology in their formation, i.e. which is the parent or the child, since the region was initially a single grain. Reaching this objective requests to go deeper in the analysis and to consider misorientations between non-neighboring HMD.

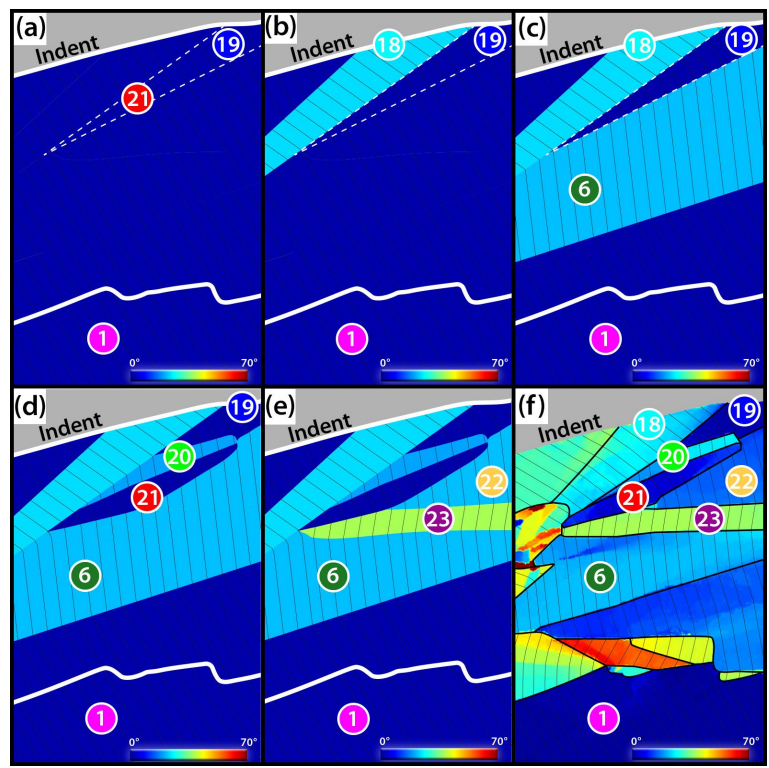

Figure 7 : (a) to (e) : schematic representation of the genesis of the twin structure described in figure 5. (f): final misorientation map established with Automated Crystal Orientation Mapping (ACOM) ASTAR.

From the previous analysis and considering the misorientation map, the following scenario, schematized in Fig. 7, can be proposed. The regions 19 and 21, defined in Fig. 3a, are both in the same crystallographic orientation as the region 1, which is the non-deformed matrix. They are assumed to originate from a same initial region (in dotted line Fig 7a) which is thus the parent and can be used as a starting point for the scenario. The boundary of the final structure is also represented in white bold line.

The HMD 18 presents a misorientation angle of 24° relative to the matrix 1 (and thus to the HMD 19 and 21). From the analysis of the pole figure (cf. fig 8), it appears that HMD 18 corresponds to a twin variant V1 $(\bar{1}2\bar{1}\bar{2})_1$ relative to the non-deformed matrix (the index 1 indicates the parent lattice used for the Miller indices). This twin is thus assumed to develop first as shown in figure 7b, creating a triangular shape twin. Then, a second twin variant V2 $(\bar{1}2\bar{1}2)_1$, relative to the non-deformed matrix 1, develops, leading to the formation of the HMD 6 (cf. Fig 7c). The misorientation angle between HMD 6 and the matrix 1 is indeed 24°, and figure 8a shows that HMD 18 and HMD 6 both present a $\{11\bar{2}2\}$ pole that overlaps with one of the matrix 1. However, the pole is different for HMD 6 and HMD 18 showing that they are two different variants. Furthermore, the misorientation between the HMD 6 and 18 is 48° which is in very good agreement with the expected misorientation angle between the variants $(\bar{1}2\bar{1}\bar{2})_1$ and $(\bar{1}2\bar{1}2)_1$ of 49.6° for $Ti_2AlN$ (cf Fig 8b). Then, the variant V2 that initially developed in HMD 6, extends through the region 19 (Fig. 7d), thus creating the HMD 20, and dividing the initially non-deformed region 19 in two regions: 19 and 21. It must be pointed out that the residual shear strain associated to this twin band is probably the reason for the motion of the twin boundary between HMD 19 and 21. This explains why this last boundary was not perfectly aligned with the expected corresponding $\{11\overline{2}2\}$ plane trace (cf. Fig. 5).

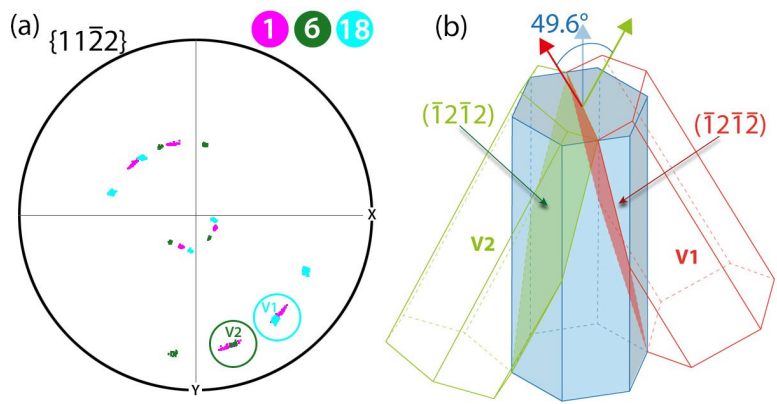

Figure 8 : a) $\{11\overline{2}2\}$ pole figure for the matrix 1 and the HMD 6 and 18. HMD 6 and 18 are two $\{11\overline{2}2\}$ twin variants relative to the parent 1. (b) Misorientation angle between the twin variants V1 ($\overline{1}2\overline{1}\overline{2}$) and V2 ($\overline{1}2\overline{1}2$).

Lastly, a secondary $\{11\overline{2}2\}$ twin appears in the HMD 6 (Fig. 7e) dividing it into HMD 6 and HMD 22 and creating the HMD 23. The HMD 21 and 23 can thus be seen as two variants $(\overline{1}2\overline{1}\overline{2})_{22}$ and $(\overline{1}2\overline{1}2)_{22}$ relative to the HMD 22. The misorientation between HMD 23 and 21 is 44° which is in rather good agreement with the 49.6° expected between these two variants in Ti$_2$AlN (cf. Fig 8b).

Fig. 7f presents the experimental misorientation map which is the real final stage. Finally, this scenario, which can account for the final structure, reinforces the hypothesis of twinning as the deformation process involved in this sample.

## 4.3.    Identification of $\{11\overline{2}1\}$ twin boundaries

A second region corresponding to HMD 7 and 8 has been investigated in a similar manner. Here again, the deformation is quite simple since these two HMD are rather isolated. Fig. 9 is a magnification of the misorientation map presented in figure 3a in the region of HMD 7 and 8, where the local misorientation across these HMD boundaries have been added inside orange arrows. The region labelled 1 is, here again, the undeformed matrix. Considering the 23° misorientation between the HMD 7 and the matrix 1, as well as the $\{11\overline{2}2\}$ pole figure and the direction of the boundary trace transferred to the pole figure, it is clear that, as for the configuration analyzed in figure 5, the boundary between the HMD 7 and the matrix is a $(\overline{1}2\overline{1}\overline{2})_1$ twin boundary. In the case of the HMD 8, the misorientation relative to the matrix is 12°. This value is very close to the misorientation of a $\{11\bar{2}1\}$ twin boundary which is 12.6° for the c/a ratio of Ti$_2$AlN. The $\{11\bar{2}1\}$ pole figure has thus been plotted for the HMD 8 and the matrix 1 in Fig. 9. This pole figure, as well as the trace orientation of the HMD 8 – matrix boundary show that this boundary is a $(\bar{1}2\bar{1}1)_1$ twin boundary. The misorientation angles between the $\{11\bar{2}1\}$ and $\{11\bar{2}2\}$ twin variants in Ti$_2$AlN has been calculated and reported in Table 1. The expected misorientation angle between a $(\bar{1}2\bar{1}2)_1$ and a $(\bar{1}2\bar{1}1)_1$ twin variants is 37.3°, which is in very good agreement with the 37° misorientation measured between HMD 7 and 8.

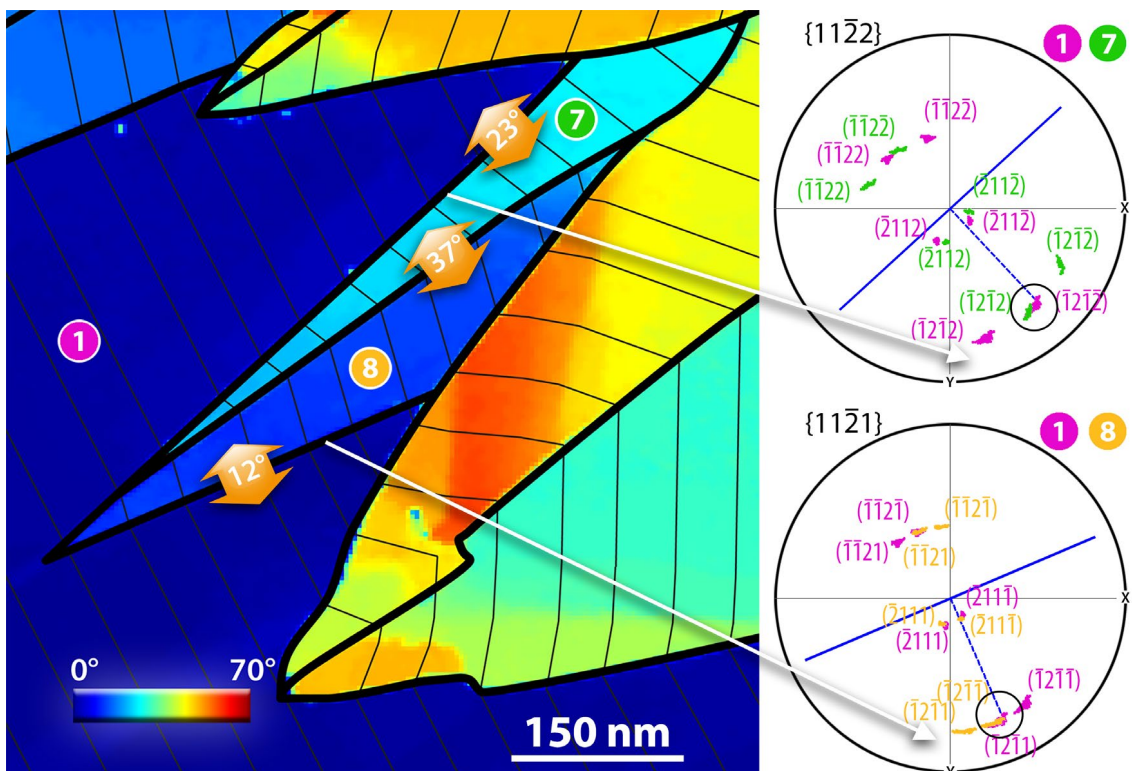

Figure 9 : Misorientation map relative to the non-deformed matrix (angular scale 0°-70°). The relative misorientation between neighboring HMD is indicated in orange arrows, and the $\{11\overline{2}2\}$ and $\{11\overline{2}1\}$ pole figures have been plotted for the HMD 7 and the matrix and for the HMD 8 and the matrix respectively. The circles in the pole figures highlight the overlapping of a pole from a HMD and the matrix, and the blue solid lines indicate the direction of the HMD boundaries transferred to the pole figure.

*Table 1: misorientation angles between $\{11\bar{2}1\}$ and $\{11\bar{2}2\}$ twin variants in Ti$_2$AlN.*

|  |  | $\{11\bar{2}1\}$ twin variants |  |  |  |  |  |
|---|---|---|---|---|---|---|---|
|  |  | $(11\bar{2}1)$ | $(\bar{1}2\bar{1}1)$ | $(\bar{2}111)$ | $(\bar{1}\bar{1}21)$ | $(1\bar{2}11)$ | $(2\bar{1}\bar{1}1)$ |
| $\{11\bar{2}2\}$ twin variants | $(11\bar{2}2)$ | 12.3° | 21.4° | 32.9° | 37.3° | 32.9° | 21.4° |
|  | $(\bar{1}2\bar{1}2)$ | 21.4° | 12.3° | 21.4° | 32.9° | 37.3° | 32.9° |
|  | $(\bar{2}112)$ | 32.9° | 21.4° | 12.3° | 21.4° | 32.9° | 37.3° |
|  | $(\bar{1}\bar{1}22)$ | 37.3° | 32.9° | 21.4° | 12.3° | 21.4° | 32.9° |
|  | $(1\bar{2}12)$ | 32.9° | **37.3°** | 32.9° | 21.4° | 12.3° | 21.4° |
|  | $(2\bar{1}\bar{1}2)$ | 21.4° | 32.9° | 37.3° | 32.9° | 21.4° | 12.3° |

# 5. Discussion about twinning in Ti$_2$AlN

Obviously, there are complex deformation mechanisms taking place in the central region below the indent. This region, along the indentation axis, is the one that has undergone the most extensive deformation, since it has accumulated all the deformation stages during the development of the indent. The above described configurations show how the activation of different twin variants, as well as secondary twinning, may lead to a complex final microstructure with many different local misorientations due to the interaction of these successive defects. Furthermore, a deformation twin that does not cross the whole grain leads to high shear strain that requires more plasticity to be relaxed. This local strain is likely to promote conventional plasticity or kink band formation [23,26–29]. This can explain the even more complex microstructure observed along the indentation axis, as well as the misorientation spreading observed in the pole figure inside the HMD (cf. fig 5a).

It is clear from the experimental results presented here that kinking and twinning are two deformation process that can coexist in the $Ti_2AlN$ MAX phase. However, even if a kink boundary can be seen as a symmetry plane in the local microstructure, this is different from a twin boundary. The main points that allows to discriminate between kink bands and twins from these experiments are the following:

1. the misorientation angles measured here are quantified and correspond to the expected twinning misorientation angle, whereas kinking must lead to many different misorientation angles depending on the dislocation density.

2. The misorientations across the tested boundaries are abrupt, whereas misorientation gradient would be expected close to a dislocation wall of high dislocation density.

3. The crystallographic relationship between two neighbor domains (cf. Fig. 6) corresponds to the exact crystallographic relationship expected for the twins tested here.

Finally, this study shows that a specific experimental approach, based on crystallographic orientation analysis is required to differentiate between kink and twin, and this approach has never been carried out on MAX phases before. This detailed and exhaustive crystallographic analysis was made possible tanks to the use of the recent ACOM ASTAR technique.

The results presented in this paper show that $\{11\overline{2}2\}$ and $\{11\overline{2}1\}$ deformation twinning can be activated, in a very significant manner, in $Ti_2AlN$ during a nanoindentation test. This is the first time that deformation twinning is reported and characterized in a MAX phase. Similar lenticular shape subgrains have been already reported by Griggs *et al.* for spherical nanoindentation at room temperature in the MAX phase $Ti_3SiC_2$ [34]. However, they did not perform crystallographic orientation analysis and concluded from their TEM images that these misoriented subgrains were kink bands. The $\{11\overline{2}2\}$ twin is not the most reported twin in hexagonal metals, but it is a compression twin: as such, it can accommodate compression stress along the c-axis. It should be recalled that the stress field generated during a Berkovich indentation test is complex, but it is mainly compressive in nature all around the indent. Furthermore, the presence of a grain boundary between grain 1 and grain 2 (cf fig 1a) is likely to play a confining role, and thus may enhance the compressive stress. Compression twins are thus consistent with the loading conditions.

As explained before, the indent studied here has been performed at high temperature, since the initial purpose was to investigate the deformation mechanisms involved in the ductile regime. However, the temperature brittle to ductile transition in $Ti_2AlN$ appears at about 900°C [36] and the temperature for the indent was only 800°C, and with a non-heated indenter tip. It is thus likely that the deformation temperature was below the brittle to ductile transition temperature. In any case, our results show that twin deformation occurs under the present experimental conditions. Further studies at lower temperature may reveal how twinning occurrence is affected by temperature and whether it is also prevalent under room temperature conditions or not.

## 6. Conclusion

A combined TEM and ACOM-ASTAR analysis of the plastic volume below a nanoindentation imprint in a single grain of $Ti_2AlN$ have been performed in order to determine the plastic deformation mechanisms involved in a MAX phase in single crystal conditions. Thanks to this experimental approach, dislocation walls, which are classically reported dislocation structures for MAX phases, have been characterized at rather large distance from the residual indent, both in terms of dislocation organization, through the TEM analysis, and in terms of low angle tilt boundary through the ACOM-ASTAR crystallographic orientation results. This approach showed the perfect complementarity between these two techniques in this kind of study. But the main result from this study comes from the analysis of highly misoriented domains (HMD) below the indent. For the less deformed region, close to the edge of the indent, these HMD have been proved to be $\{11\overline{2}2\}$ deformation twins. This identification is based on the crystallographic analysis method proposed by [38,39], and is reinforced both by the simulation of the full stereographic projections, and by the proposal of a complete twinning based scenario fully consistent with the final microstructure. Two variants have been identified as well as secondary twinning. $\{11\overline{2}2\}$ and $\{11\overline{2}1\}$ twins have also been identified forming two needle shape neighboring domains. Further studies are now required to understand the deformation mechanisms in the more deformed region, but the results presented here suggest that the HMD along the indentation axis are the results of successive twinning events interacting with dislocation walls.

This study, which has revealed and characterized for the first time deformation twinning in a MAX phase, shed a new light on the understanding of the mechanical behavior of this class of material. The role of temperature should now be investigated through experiments at ambient conditions. Further studies, using a spherical indenter instead of a Berkovich one, may also be considered in order to generate less dense structures and to understand how deformation twinning interacts with basal slip in the deformation process of MAX phases.

# References

[1]    W. Jeitschko, H. Holleck, H. Nowotny, F. Benesovsky, Die Verbindungen RuGa und RuGa2, Monatshefte Fur Chemie. 94 (1963) 838–840.

[2]  V.H. Nowotny, Strukturchemie einiger Verbindungen der U?bergangsmetale mit den elementen C, Si, Ge, Sn, Progress in Solid State Chemistry. 5 (1971) 27–70.

[3]  M.W. Barsoum, T. El-Raghy, Synthesis and characterization of a remarkable ceramic: $Ti_3SiC_2$, Journal of the American Ceramic Society. 79 (1996) 1953–1956.

[4]  M.W. Barsoum, $M_{n+1}AX_n$ phases: a new class of solids thermodynamically stable nanolaminates, Progress in Solid State Chemistry. 28 (2000) 201–281.

[5]  M.W. Barsoum, M. Radovic, Mechanical Properties of the MAX Phases, in: Encyclopedia of Materials: Science and Technology, Elsevier, 2004: pp. 1–16.

[6]  M.W. Barsoum, T. El-Raghy, The MAX Phases: Unique New Carbide and Nitride Materials: Ternary ceramics turn out to be surprisingly soft and machinable, yet also heat-tolerant, strong and lightweight, American Scientist. 89 (2001) 334–343.

[7]  M.W. Barsoum, T. El-Raghy, Room-temperature, ductile carbides, Metallurgical and Materials Transactions A: Physical Metallurgy and Materials Science. 30 (1999) 363–369.

[8]  A. Guitton, A. Joulain, L. Thilly, C. Tromas, Dislocation analysis of $Ti_2AlN$ deformed at room temperature under confining pressure, Phil. Mag. 92 (2012) 4536–4546.

[9]  L. Farber, I. Levin, M.W. Barsoum, HRTEM study of a low angle boundary in plastically deformed $Ti_3SiC_2$, Phil Mag Lett. 79 (1999) 163.

[10]  B.J. Kooi, R.J. Poppen, N.J.M. Carvalho, J.T.M. De Hosson, M.W. Barsoum, $Ti_3SiC_2$: A damage tolerant ceramic studied with nano-indentations and transmission electron microscopy, Acta Materialia. 51 (2003) 2859–2872.

[11]  M.W. Barsoum, M. Radovic, Elastic and Mechanical Properties of the MAX Phases, Annu. Rev. Mater. Res. 41 (2011) 195–227.

[12]  A. Joulain, L. Thilly, J. Rabier, Revisiting the defect structure of MAX phases: The case of $Ti_4AlN_3$, Philosophical Magazine. 88 (2008) 1307–1320.

[13]  E. Drouelle, A. Joulain, J. Cormier, V. Gauthier-Brunet, P. Villechaise, S. Dubois, P. Sallot, Deformation mechanisms during high temperature tensile creep of $Ti_3AlC_2$ MAX phase, Journal of Alloys and Compounds. 693 (2017) 622-622–630.

[14]  X.H. Wang, Y.C. Zhou, Microstructure and properties of Ti3AlC2 prepared by the solid–liquid reaction synthesis and simultaneous in-situ hot pressing process, Acta Materialia. 50 (2002) 3143–3151.

[15]  M.W. Barsoum, L. Farber, T. El-Raghy, Dislocations, kink bands, and room-temperature plasticity of $Ti_3SiC_2$, Metallurgical and Materials Transactions A: Physical Metallurgy and Materials Science. 30 (1999) 1727–1738.

[16]  Y. Wada, N. Sekido, T. Ohmura, K. Yoshimi, Deformation Microstructure Developed by Nanoindentation of a MAX Phase $Ti_2AlC$, Mater. Trans. 59 (2018) 771–778.

[17]  E. Orowan, Nature, A Type of Plastic Deformation New in Metals. 149 (1942) 463.

[18]  J.B. Hess, C.S. Barrett, Structure and nature of kink bands in zinc, Trans. AIME. 185 (1949) 599–606.

[19]  A. Serra, D.J. Bacon, Computer simulation of screw dislocation interactions with twin boundaries in H.C.P. metals, Acta Metallurgica et Materialia. 43 (1995) 4465–4481.

[20]  J.W. Christian, Deformation Twinning, in: The Theory of Transformations in Metals and Alloys, Elsevier, 2002: pp. 859–960.

[21]  A. Kelly, K.M. Knowles, Crystallography and crystal defects, Third edition, Wiley, Hoboken, NJ, 2020.

[22]  P.G. Partridge, The crystallography and deformation modes of hexagonal close-packed metals, Metal. Rev. 12 (1967) 169–194.

[23]  J.W. Christian, S. Mahajan, Deformation twinning, Prog. Mat. Sci. 39 (1995) 1–157.

[24]  R.J.D. Tilley, Understanding solids: the science of materials, J. Wiley, Chichester, West Sussex, England ; Hoboken, NJ, USA, 2004.

[25] T. El-Raghy, M.W. Barsoum, A. Zavaliangos, S.R. Kalidindi, Processing and mechanical properties of $Ti_3SiC_2$: II, Effect of grain size and deformation temperature, Journal of the American Ceramic Society. 82 (1999) 2855–2860.

[26] D.C. Jillson, An experimental survey of deformation and annealing processes in zinc, JOM. 2 (1950) 1009–1018.

[27] A.J.W. Moore, Twinning and accommodation kinking in zinc, Acta Metallurgica. 3 (1955) 163–169.

[28] C. Zambaldi, C. Zehnder, D. Raabe, Orientation dependent deformation by slip and twinning in magnesium during single crystal indentation, Acta Materialia. 91 (2015) 267–288.

[29] E. Roberts, P.G. Partridge, The accommodation around $\{10\bar{1}2\}$ ⟨$10\bar{1}1$⟩ twins in magnesium, Acta Metallurgica. 14 (1966) 513–527.

[30] H. Zhang, T. Hu, X. Wang, Y. Zhou, Structural defects in MAX phases and their derivative MXenes: A look forward, Journal of Materials Science & Technology. 38 (2020) 205–220.

[31] J.M. Molina-Aldareguia, J. Emmerlich, J.P. Palmquist, U. Jansson, L. Hultman, Kink formation around indents in laminated $Ti_3SiC_2$ thin films studied in the nanoscale, Scripta Materialia. 49 (2003) 155–160.

[32] C. Tromas, P. Villechaise, V. Gauthier-Brunet, S. Dubois, Plasticity of the new MAX phase $Ti_3SnC_2$ studied by nanoindentation, (2009).

[33] Y. Kabiri, N. Schrenker, J. Müller, M. Mačković, E. Spiecker, Direct observation of dislocation formation and plastic anisotropy in $Nb_2AlC$ MAX phase using in situ nanomechanics in transmission electron microscopy, Scripta Materialia. 137 (2017) 104–108.

[34] J. Griggs, A.C. Lang, J. Gruber, G.J. Tucker, M.L. Taheri, M.W. Barsoum, Spherical nanoindentation, modeling and transmission electron microscopy evidence for ripplocations in $Ti_3SiC_2$, Acta Materialia. 131 (2017) 141–155.

[35] J.S.K.-L. Gibson, S. Schröders, C. Zehnder, S. Korte-Kerzel, On extracting mechanical properties from nanoindentation at temperatures up to 1000 °C, Extreme Mechanics Letters. 17 (2017) 43–49.

[36] A. Guitton, A. Joulain, L. Thilly, C. Tromas, Evidence of dislocation cross-slip in MAX phase deformed at high temperature, Scientific Reports. 4 (2014).

[37] I. Horcas, R. Fernandez, J.M. Gomez-Rodriguez, J. Colchero, J. Gomez-Herrero, A.M. Baro, WSXM: A software for scanning probe microscopy and a tool for nanotechnology, Review of Scientific Instruments. 78 (2007) 013705.

[38] S.I. Wright, R.J. Larsen, Extracting twins from orientation imaging microscopy scan data., Journal of Microscopy. 205 (2002) 245-245–252.

[39] T.A. Mason, J.F. Bingert, G.C. Kaschner, S.I. Wright, R.J. Larsen, Advances in deformation twin characterization using electron backscattered diffraction data, in: 2002: pp. 949-949–54.

[40] P.E. Marshall, G. Proust, J.T. Rogers, R.J. Mccabe, Automatic twin statistics from electron backscattered diffraction data: AUTOMATIC TWIN STATISTICS FROM EBSD DATA, Journal of Microscopy. 238 (2009) 218–229.

[41] N. Bozzolo, L. Chan, A.D. Rollett, Misorientations induced by deformation twinning in titanium, Journal of Applied Crystallography. 43 (2010) 596-596–602.

[42] Y.J. Chen, Y.J. Li, X.J. Xu, J. Hjelen, H.J. Roven, Novel deformation structures of pure titanium induced by room temperature equal channel angular pressing, Materials Letters. 117 (2014) 195-195–8.

[43]  S. Xu, M. Gong, X. Xie, Y. Liu, C. Schuman, J.-S. Lecomte, J. Wang, Crystallographic characters of {11$\bar{2}$2} twin-twin junctions in titanium., Philosophical Magazine Letters. 97 (2017) 429-429–441.

[44]  S. Xu, M. Gong, Y. Jiang, C. Schuman, J.-S. Lecomte, J. Wang, Secondary twin variant selection in four types of double twins in titanium, Acta Materialia. 152 (2018) 58-58–76.

[45]  B. Beausir, J.-J. Fundenberger, ATOM - Analysis Tools for Orientation Maps, http://atom-software.eu/, Université de Lorraine - Metz, 2015.

[46]  R. Krakow, R.J. Bennett, D.N. Johnstone, Z. Vukmanovic, W. Solano-Alvarez, S.J. Lainé, J.F. Einsle, P.A. Midgley, C.M.F. Rae, R. Hielscher, On three-dimensional misorientation spaces, Proc. R. Soc. A. 473 (2017) 20170274.

[47]  F. Mompiou, R. Xie, *pycotem* : An open source toolbox for online crystal defect characterization from TEM imaging and diffraction, Journal of Microscopy. (2020) jmi.12982.

[48]  B.A. Bilby, A.G. Crocker, A.H. Cottrell, The theory of the crystallography of deformation twinning, Proceedings of the Royal Society of London. Series A. Mathematical and Physical Sciences. 288 (1965) 240–255.

[49]  M. Niewczas, Lattice correspondence during twinning in hexagonal close-packed crystals, Acta Materialia. 58 (2010) 5848–5857.

[50]  T.B. Britton, J. Jiang, Y. Guo, A. Vilalta-Clemente, D. Wallis, L.N. Hansen, A. Winkelmann, A.J. Wilkinson, Tutorial: Crystal orientations and EBSD — Or which way is up?, Materials Characterization. 117 (2016) 113–126.

[51]  M.V. Klassen-Neklyudova, Mechanical Twinning of Crystals, Springer US, Boston, MA, 1964.